\documentclass[twocolumn,showpacs,preprintnumbers,amsmath,amssymb,floatfix]{revtex4}
\usepackage{graphicx}

\begin{document}

\title{Doping dependent charge injection and band alignment in organic field-effect transistors}

\author{B.\ H.\ Hamadani$^1$, H.\ Ding$^2$, Y.\ Gao$^2$, and D.\ Natelson$^{1,3}$}

\affiliation{$^1$ Department of Physics and Astronomy, Rice University, 6100 Main St., Houston, TX 77005}

\affiliation{$^2$ Department of Physics and Astronomy, University
of Rochester, Rochester, NY 14627}

\affiliation{$^3$ Department of Electrical and Computer
Engineering, Rice University, 6100 Main St., Houston, TX 77005}

\pacs{73.30.+y,73.61.Jc,73.61.Ph}

\begin{abstract}
We have studied metal/organic semiconductor charge injection in
poly(3-hexylthiophene) (P3HT) field-effect transistors with Pt and Au
electrodes as a function of annealing in vacuum.  At low impurity
dopant densities, Au/P3HT contact resistances increase and become
nonohmic.  In contrast, Pt/P3HT contacts remain ohmic even at far
lower doping.  Ultraviolet photoemission spectroscopy (UPS) reveals
that metal/P3HT band alignment shifts dramatically as samples are
dedoped, leading to an increased injection barrier for holes, with a
greater shift for Au/P3HT.  These results demonstrate that doping can
drastically alter band alignment and the charge injection process at
metal/organic interfaces.
\end{abstract}

\maketitle

Charge injection from metals into disordered organic semiconductors
(OSCs) is a complicated physical process for which no complete unified
picture exists.  Charge transport in disordered OSCs is by hopping
through a density of localized states that depends strongly on energy,
so that the local chemical potential has profound effects on charge
mobility\cite{Vissenberg,Pasveer}.  Field-effect transistors (FETs)
may be used as tools to examine injection, by differentiating between
the contact resistance, $R_{c}$, and the intrinsic channel resistance,
$R_{ch}$.  Approaches include analyses of single device
characteristics\cite{Horowitz,Street}, scanning
potentiometry\cite{Seshadri,BurgiAPL,BurgiJAP}, gated four-probe
measurements\cite{Gate4probe}, and scaling of total device resistance
with channel length in a series of
devices\cite{Klauk,Necludiov,OurAPL,Zaumseil,Meijer,Blanchet}.  Models
of injection adapted from inorganic semiconductors have had mixed
results.  For ohmic injection between Au and P3HT\cite{OurAPL}, the
contact resistivity is inversely proportional to the intrinsic channel
mobility over 4 decades, over a broad range of temperatures and gate
voltages, consistent with diffusion-limited
injection\cite{Mallia1,Mallia2}.  However, when a significant barrier,
$\Delta$ exists between the metal Fermi level and the OSC valence band
(Cr/P3HT or Cu/P3HT), thermionic emission models cannot explain the
field and temperature dependence of the injected
current\cite{BurgiJAP,ourJAP2}. A more sophisticated
model~\cite{Arkhipov,Arkhipov2} incorporating a hopping injection into
a disordered density of localized states with emphasis on the primary
injection event\cite{ourJAP2,VanWou} explains this nonohmic charge
injection data more consistently.

Dopant density strongly influences the magnitude and mechanism of
charge injection into OSCs.  A doping-dependent charge injection
study\cite{Rep} into P3HT using planar and sandwich geometries has
indicated that there are severe contact limitations at low doping
densities.  Hosseini et.\ al.\ have also shown~\cite{Hosseini} that
contact resistance in disordered OSCs significantly decreases at high
doping concentrations due to dopant-induced broadening of the Gaussian
density of localized states.

In this paper, we examine doping dependent charge injection in a
series of bottom contact field effect transistors using Au and Pt as
the contacting electrodes and P3HT as the active organic polymer, and
find an additional effect of doping.  From the length dependence of
the total device resistance, $R_{on}$, we extract the contact and
channel resistances as a function of the gate voltage, $V_{G}$, and
doping.  Exposure to air and humidity is known to enhance hole doping
in P3HT\cite{Hoshino}.  As we reduce the concentration of such dopants
(related to the bulk P3HT conductivity) by annealing devices in vacuum
at elevated temperatures, $R_{c}$ and the ratio $R_{c}/R_{ch}$
increases dramatically for Au/P3HT devices, with injection becoming
nonohmic.  For Pt/P3HT devices, $R_{c}$ remains relatively low
compared to $R_{ch}$, and injection remains ohmic even when bulk
conductivity is reduced below measurable limits.  Ultraviolet
photoemission spectroscopy (UPS) reveals that changing dopant
concentration strongly alters the band alignment between the metal
Fermi level and the OSC valence band.  As dopants are removed, the
energetic difference between the Au Fermi level and the P3HT valence
band increases by about 0.5~eV, while this effect is much less severe
in Pt/P3HT.  These results demonstrate that doping effects on surface
dipole formation, charge transfer at the metal/organic interface, and
band bending must be considered in any full treatment of metal/OSC
charge injection.

Devices are made in a bottom-contact configuration\cite{OurAPL} on a
degenerately doped $p+$ silicon wafer used as a gate. The gate
dielectric is 200 nm of thermal SiO$_{2}$.  Source and drain
electrodes are patterned using electron beam lithography in the form
of an interdigitated set of electrodes with systematic increase in the
distance between each pair. The channel width, $W$, is kept fixed for
all devices at 200~$\mu$m. The electrodes are deposited by electron
beam evaporation of 2.5~nm of Ti and 25~nm of Au or Pt followed by
liftoff. This thickness of metal is sufficient to guarantee film
continuity and good metallic conduction while attempting to minimize
disruptions of the surface topography that could adversely affect
polymer morphology.

Prior to OSC deposition, the substrates were cleaned for about 2
minutes in an oxygen plasma. The organic semiconductor, 98\%
regio-regular P3HT~\cite{Aldrich}, is dissolved in chloroform at a
0.06\% weight concentration, passed through polytetrafluoroethylene
(PTFE) 0.2\ $\mu$m filters and solution cast onto the clean substrate,
with the solvent allowed to evaporate in ambient conditions. The
resulting films are tens of nm thick as determined by atomic force
microscopy. The measurements are performed in vacuum (~$10^{-6}$ Torr)
in a variable temperature range probe station using a semiconductor
parameter analyzer.  To produce a certain doping level, the sample is
annealed at elevated temperatures ($\sim$350-380~K) in vacuum for
several hours and then cooled down to room temperature for
measurement.  We characterize the doping by the effective
conductivity, calculated from the low $V_{D}$ source-drain conductance
and the estimated P3HT layer thickness at zero gate voltage.  In the
absence of band bending at the P3HT/SiO$_{2}$ interface, there should
be no ``channel'' at the interface, and the measured source-drain
conductance should be a bulk effect.  We note that similar
conductances are found in two-terminal planar devices fabricated on
glass substrates.  The conductivity after an annealing step remains
stable in vacuum at room temperature and below on the timescale of the
measurements, indicating no further change in doping.

\begin{figure}[h!]
\begin{center}
\includegraphics[clip, width=8cm]{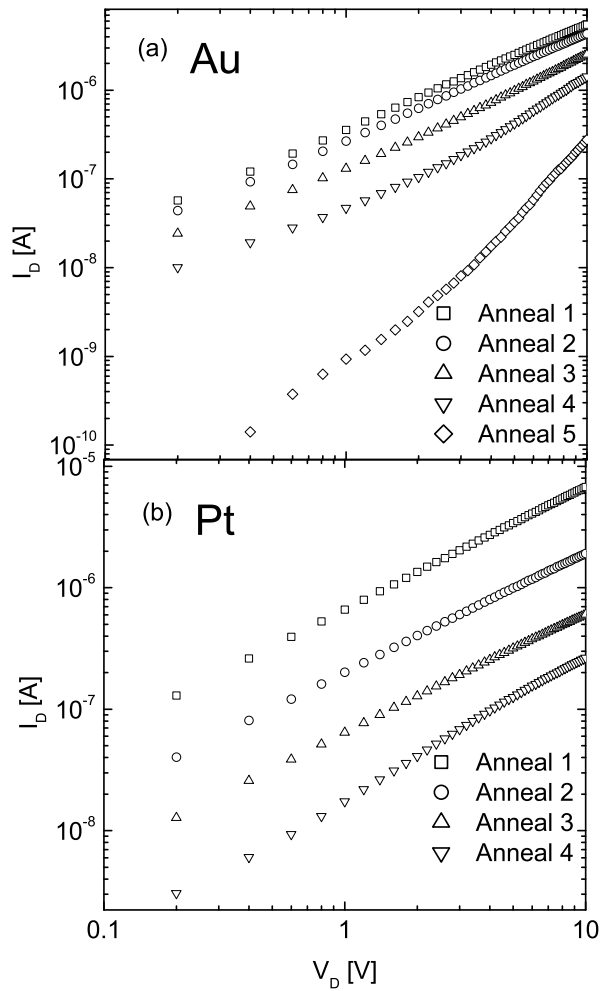}
\end{center}
\vspace{-3mm}
\caption{(a) Log-Log plot of the transport characteristics of a
Au/P3HT device with $L=10~\mu$m at $T=300$~K and at a fixed
$V_{G}=-60$~V for different annealing as described in the text.
(b) Similar plot for a Pt/P3HT device with identical geometry as
(a).}
\label{fig1}
\end{figure}

The devices operate as standard p-type FETs in accumulation
mode\cite{OurAPL}.  With source electrode grounded, the devices are
measured in the shallow channel regime ($V_{D}< V_{G}$).
Figure~\ref{fig1}a shows the transport characteristics of a Au/P3HT
device with $L=10$\ $\mu$m at $T=300$~K and at a fixed $V_{G}=-60$ V
for different doping levels.  Table~\ref{tab1} shows the annealing
schedule for the Au/P3HT and Pt/P3HT devices.  After the fourth
annealing step source-drain transport in the Au/P3HT devices was
highly nonlinear.  In contrast, Fig.~\ref{fig1}b shows a similar plot
for a Pt/P3HT device with identical parameters as Au above.  After a
more extensive annealing process such that bulk P3HT conductivity was
reduced below measurable limits, the $I_{D}-V_{D}$ data remain nearly
linear.  To ascertain whether the annealing process irreversibly
alters the polymer or the interface, Au/P3HT samples exposed to
ambient air were re-examined after several days, and the conductivity
had returned to its initial pre-annealing levels. We carried out an
additional annealing stage at this point (A.5$^{*}$), which reproduced
the nonlinearity trend observed earlier.

\begin{table}
\caption{Annealing times and temperatures and resulting bulk P3HT
conductivities (for average film thickness of 25~nm).  For Pt
devices bulk conduction following the third and fourth anneals was
below detectable limits. A.5$^{*}$ was performed after a few days
of air exposure.}
\begin{tabular}{||c|c|c|c|c||}
  \hline
  \hline
     & stage & time\ (hrs) & T\ (K) & $\sigma$\ (S/m) \\
  \cline{2-5}
  Au & anneal\ (A).1 & 17& 350 & 0.032
 \\
   & A.2 & +7 & 350 & 0.022
 \\
   & A.3 & +12 & 350 & 0.0077
 \\
   & A.4 & +22& 350 & $8.9\times10^{-4}$ \\
   & A.5$^{*}$ & 16 & 370 & - \\
  \hline
  \hline
  Pt & A.1& 24 & 350 & 0.064
 \\
   & A.2& +17 & 360 & 0.0036
 \\
   & A.3 & +22 & 370 & - \\
   & A.4& +22 & 380 & - \\
  \hline
\end{tabular}
\label{tab1}
\end{table}

\begin{figure}[h!]
\begin{center}
\includegraphics[clip, width=8cm]{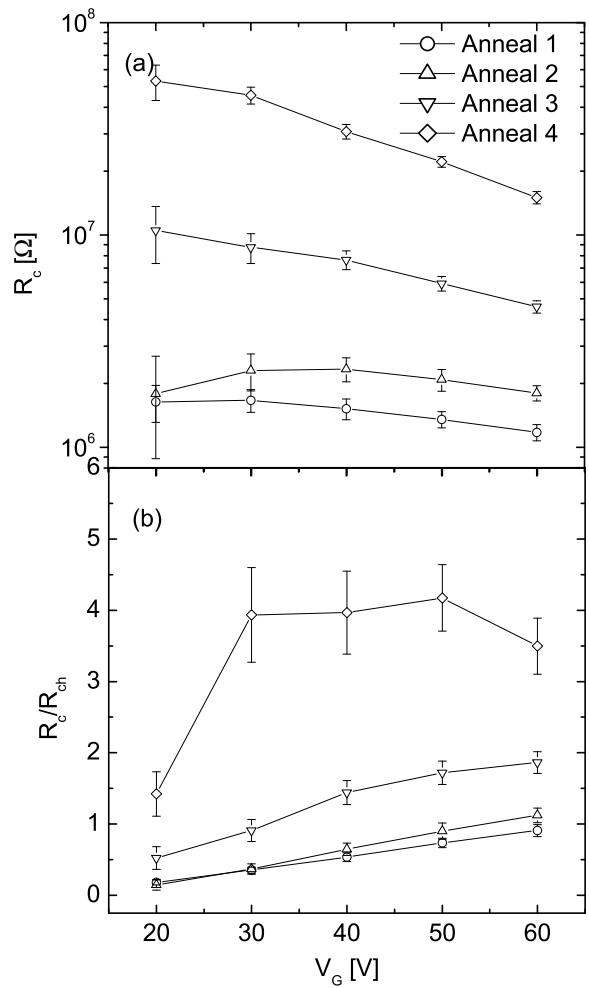}
\end{center}
\vspace{-3mm}
\caption{(a) Gate voltage dependence of $R_{c}$ for different
anneals for the Au device at room temperature. (b)
$R_{c}/R_{ch}$ as a function of $V_{G}$ for a Au device with
$L=10~\mu$m.}
\label{fig2}
\end{figure}

From the data in Fig.~\ref{fig1}, we extracted $R_{ch}$ and $R_{c}$
from the $L$ dependence of the total device resistance, $R_{on}\equiv
\partial V_{D}/\partial I_{D}$ as described in Ref~\cite{OurAPL}.
Figure~\ref{fig2}a shows the $V_{G}$ dependence of $R_{c}$ for
different annealing steps for the Au device at room temperature.  Here
we obtain $R_{c}$ in the limit $|V_{D}|< 1$~V, where transport is
still reasonably linear even at higher dedopings.  We note that we
have developed a procedure for extracting contact current-voltage
characteristics even in the limit of strong injection
nonlinearities\cite{ourJAP2}, but it is difficult to quantify such
injection by a single number such as $R_{c}$.  After each anneal, the
contact resistance increases significantly.  To test for
contact-limited transport, we plot the ratio $R_{c}/R_{ch}$ as a
function of $V_{G}$ for a Au device with $L=10$\ $\mu$m in
Fig.~\ref{fig2}b.  At higher dedopings and higher gate voltages, the
devices are clearly contact limited.  Thus the nonlinear transport
seen in this regime at higher biases indicates the possible formation
of a charge injection barrier for holes.  Analogous data for Pt
devices (not shown) reveals that the contact resistance is lower than
that that observed for Au, and that $R_{c}/R_{ch}$ remains below 1,
only reaching approximately 1 at the most severely dedoped levels.

\begin{figure}[h!]
\begin{center}
\includegraphics[clip, width=8cm]{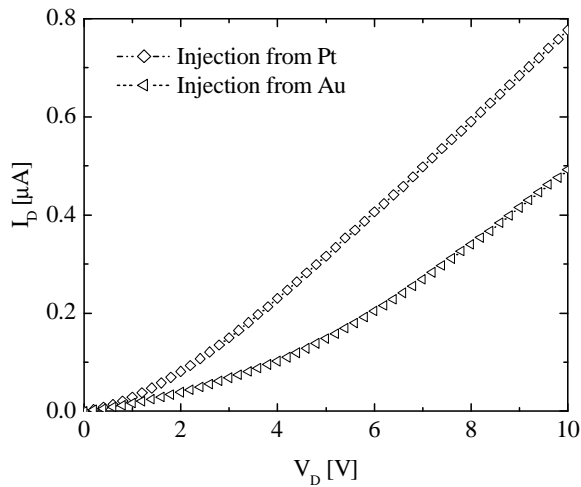}
\end{center}
\vspace{-3mm} 
\caption{Linear plot of $I_{D}-V_{D}$ for injection from
Au and Pt for a sample at 300~K and $V_{G}=-80$~V.  This sample was
dedoped such that bulk conductivity at 300~K was below our
measurement threshold.}
\label{fig3}
\end{figure}

To further confirm that charge injection from Au electrodes becomes
more difficult at lower doping levels than injection from Pt, we
fabricated a series of devices (in a two-step lithography process)
with \emph{alternating} Au and Pt electrodes. The data are taken twice
for each device, once with the source electrode on Au with the drain
on Pt and the second time vice versa.  At higher doping levels, the
transport data are similar for injection from Au and Pt, but as the
sample is annealed, injection from Au becomes more nonlinear and
allows for less current at low drain biases.  Figure~\ref{fig3} shows
a linear plot of $I_{D}-V_{D}$ for injection from Au and Pt for a
certain dedoped level at $T= 300 K$ and $V_{G}= -80 V$ with $L=7$\
$\mu$$m$ and $W=200$\ $\mu$$m$.  Previous scanning potentiometry
experiments~\cite{Seshadri,BurgiAPL,BurgiJAP} reveal that in systems
with significant $\Delta$ most of the potential drop due to contacts
occurs at the source.  Fig.~\ref{fig3} is consistent with the the
formation of a larger injection barrier between Au and P3HT than Pt
under identical annealing conditions.  Note that the lack of
nonlinearity in Pt/P3HT/Pt (hole injection from Pt, hole collection by
Pt) devices (see Fig.~\ref{fig1}b) compared with the nonlinear data of
Fig.~\ref{fig3} in the Au/P3HT/Pt (hole injection from Au, hole
collection by Pt) configuration further supports the conclusion that
the {\it injecting} contact is the source of the nonlinearities.

Dopant concentration clearly affects charge transport across the
interface between the OSC and injecting electrodes. Therefore
examining the band alignment and interfacial dipole is
imperative. Ultraviolet photoelectron spectroscopy (UPS) is a useful
tool to monitor changes in valance electronic structure and
workfunction of the interface. Previous studies of pentacene-based
devices\cite{Gau} have shown interface dipole formation at the
metal/organic interface which varies linearly with the measured metal
workfunction. Here, we show the results of our UPS study for Au/P3HT
and Pt/P3HT interfaces.  Details of the UPS setup can be found in
Ref.~\cite{Gau}.  Samples were prepared by solution casting of P3HT on
thin films (25 nm) of Au or Pt using the same procedures as in
the FET devices.

\begin{figure}[h!]
\begin{center}
\includegraphics[clip, width=8cm]{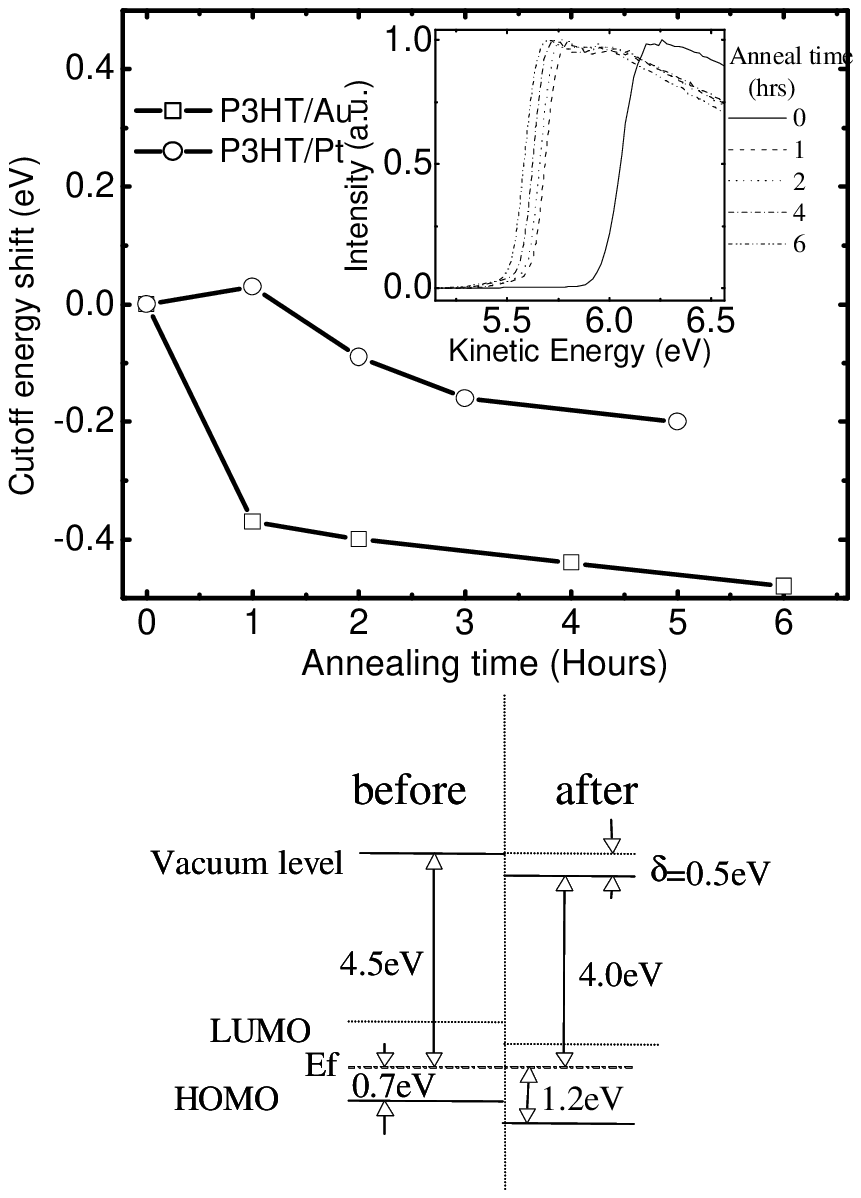}
\end{center}
\vspace{-3mm}
\caption{(a) UPS cutoff energy shift as a function of annealing time for
both P3HT/Au and P3HT/Pt. Inset: UPS cutoff of P3HT/Au sample for
different annealing times at $350 K$ (b) Energy level diagram 
of band alignment, based on the results of the UPS data for Au/P3HT
before and after the annealing process, showing the large change
in the barrier for hole injection.}
\label{fig4}
\end{figure}

The inset in Figure~\ref{fig4}a shows the UPS cutoff of a P3HT/Au
sample for different annealing times at 350~K. After 6 hours of
annealing, the total change in the vacuum level, which appears as a
series of shifts in the cutoff data, is about 0.5~eV. The main part of
Fig.~\ref{fig4}a plots the energy shift as a function of annealing
time for both Au/P3HT and Pt/P3HT. The cutoff shift for Pt/P3HT after
5 hours is 0.2~eV.  These shifts correspond to an increase in the
Au/P3HT bulk valence level by these amounts.  Fig.~\ref{fig4}b shows
the band alignment inferred from UPS for Au/P3HT before and after the
annealing process.  The appearance of 0.5~eV of additional hole
injection barrier is consistent with increased contact resistance and
nonlinear charge injection data shown in Fig.~\ref{fig1}a.  Since the
energy shift for Pt/P3HT samples is smaller, less contact limitations
are expected and observed for these devices.  The difference between
Au and Pt is likely related to their differing work functions and
surface chemistries.

The UPS measurements and energy level diagrams of Fig.~\ref{fig4} do
not directly probe the band alignment at the metal/organic interface,
as the solution-cast P3HT layer is too thick to permit direct
assessment of the metal/P3HT interface.  Two effects difficult
to discriminate in these samples may contribute to the changes in
level alignment and injection mechanism with doping:  band bending
in the bulk, and surface dipole modification directly at the interface.

Band bending effects have been seen in an experiment\cite{Kahn}
involving $p$-doping of zinc phthalocyanine, where Gao and Kahn have
shown that in addition to the reduction of the interface dipole upon
doping, the valence (highest occupied molecular orbital, HOMO) level
shifts towards $E_{F}$ within a layer thickness of a few nanometers.
This shift indicates the formation of a space charge region and band
bending near the interface.  The improved transport at high doping
levels in that experiment is associated with an increase in film
conductivity and tunnelling of carriers through the now-thin interface
barrier.  

The other possible contribution to the observed change in the
injection barrier by annealing is the removal of the interface
impurities, most likely H$_{2}$O.  Heating inside vacuum can remove
H$_{2}$O from the metal electrode surface, resulting in more direct
contact of the organic with the electrode. The interface barrier in
this case will be determined by the metal/organic interface dipole,
and it will be smaller for Pt than for Au because of the higher
workfunction of Pt\cite{Gau}. The smaller barrier for Pt is also
supported by the less change of the cutoff upon annealing as observed
with UPS.  Although the formation of a direct contact to metal will
increase the barrier for both Pt and Au, the smaller value of the
increase still allows Pt contact to be ohmic.  Once re-exposed to the
ambient, impurities may diffuse back into the interface, resulting in
a recovery of the injection properties of the un-annealed devices. The
reduction of the hole injection barrier by ambient exposure of the
metal electrode has recently been observed\cite{Wan} in another OSC on
Au by Wan et al.  

We note, however, that the annealing processes used
in this work are very mild compared to those typically used to remove
physisorbed interfacial impurities such as water.  Typical UHV baking
procedures for desorption require temperatures in excess of 370~K,
while we observe significant effects even at 320~K.  This 
suggests that desorption of interfacial impurities is unlikely
to be the dominant source of the observed effects.


To examine the physics of charge injection from metals into disordered
organic semiconductors, we examined transport properties of a series
of organic FETs with P3HT as the active polymer layer and Pt and Au as
the source/drain electrodes as a function of annealing and resulting
dopant concentration.  We extract the contact and channel resistances
from the length dependence of the resistance, and observed that the
contact resistance is dominant and strongly nonlinear at lower dopant
concentrations and higher gate voltages for Au/P3HT samples.  These
effects are much less severe in Pt/P3HT samples.  UPS data reveal that
upon dedoping, the energy levels shift at the interface, leading to an
increased barrier for hole injection. This shift is stronger for Au
samples than Pt, consistent with strong nonlinear charge injection
observed for Au samples at high dedopings.  These results demonstrate
that doping can profoundly affect the physics of charge injection in
such systems by strongly altering the band alignment between the metal
and the organic.  The scale of the interface dipole shifts can
significantly exceed the dopant-induced broadening of the density of
states.  Understanding such interfacial charge transfer and band
alignment is essential to a complete description of metal/OSC
interfaces.

DN and BH acknowledge partial support by the Research Corporation
and the David and Lucille Packard Foundation.  HD and YG 
acknowledge support in part by NSF DMR-0305111.





\begin{thebibliography}{99}
\bibitem{Vissenberg}
M.C.J.M. Vissenberg and M. Matters, Phys. Rev. B \textbf{57}, 12964 (1998).
\bibitem{Pasveer}
W.F. Pasveer, J. Cottaar, C. Tanase, R. Coehoorn, P.A. Bobbert,
P.W.M. Blom, D.M. de Leeuw, and M.A.J. Michels,
Phys. Rev. Lett. \textbf{94}, 206601 (2005).
\bibitem{Horowitz} G. Horowitz, R. Hajlaoui, D. Fichou, and A. El
Kassmi, J. Appl. Phys.\textbf{85}, 3202 (1999).
\bibitem{Street} R. A. Street and A. Salleo, Appl. Phys. Lett.
\textbf{81}, 2887\ (2002).
\bibitem{Seshadri} K. Seshadri and C. D. Frisbie, Appl. Phys.
Lett. \textbf{78}, 993\ (2001).
\bibitem{BurgiAPL} L. B\"{u}rgi, H. Sirringhaus, and R. H. Friend,
Appl. Phys. Lett. \textbf{80}, 2913\ (2002).
\bibitem{BurgiJAP} L. B\"{u}rgi, T. J. Richards, R. H. Friend, and H.
Sirringhaus, J. Appl. Phys. \textbf{94}, 6129\ (2003).
\bibitem{Gate4probe} P. V. Pesavento, R. J. Chesterfield, C. R.
Newman, and C. D. Frisbie, J. Appl. Phys. \textbf{96}, 7312\
(2004).
\bibitem{Klauk} H. Klauk, G. Schmid, W. Radlik, W. Weber, L. Zhou,
C. D. Sheraw, J. A. Nichols, and T. N. Jackson, Solid-State
Electron. \textbf{47}, 297\ (2003).
\bibitem{Necludiov} P. V. Necludiov, M. S. Shur, D. J. Gundlach,
and T. N. Jackson, Solid-State Electron. \textbf{47}, 259\ (2003)
\bibitem{OurAPL} B. H. Hamadani and D. Natelson, Appl. Phys. Lett.
\textbf{84}, 443\ (2004).
\bibitem{Zaumseil} J. Zaumseil, K. W. Baldwin, and J. A. Rogers,
J. Appl. Phys. \textbf{93}, 6117\ (2003).
\bibitem{Meijer} E. J. Meijer, G. H. Gelinck, E. van Veenendaal,
B. -H. Huisman, D. M. de Leeuw, and T. M. Klapwijk, Appl. Phys.
Lett. \textbf{82}, 4576\ (2003).
\bibitem{Blanchet} G. B. Blanchet, C. R. Fincher, and M.
Lefenfeld, Appl. Phys. Lett. \textbf{84}, 296\ (2004).
\bibitem{Mallia1} J. Campbell Scott and G. G. Malliaras, Chem. Phys.
Lett. \textbf{299}, 115\ (1999).
\bibitem{Mallia2} Y. Shen, M. W. Klein, D. B. Jacobs, J. Campbell Scott,
and G. G. Malliaras, Phys. Rev. Lett. \textbf{86}, 3867\ (2001).
\bibitem{Arkhipov} V. I. Arkhipov, E. V. Emelianova, Y. H. Tak,
and H. B\"{a}ssler, J. Appl. Phys. \textbf{84}, 848\ (1998).
\bibitem{Arkhipov2} V. I. Arkhipov, U. Wolf, and H. B\"{a}ssler,
Phys. Rev. B \textbf{59}, 7514\ (1999).
\bibitem{ourJAP2} B. H. Hamadani and D. Natelson, J. Appl. Phys.
\textbf{97}, 064508\ (2005).
\bibitem{VanWou} T. van Woudenbergh, P. W. M. Blom, M. C. J. M.
Vissenberg, and J. N. Huiberts, Appl. Phys. Lett. \textbf{79},
1697\ (2001).
\bibitem{Hosseini} A. R. Hosseini, M. H. Wong, Y. Shen and G. G.
Malliaras, J. Appl. Phys. \textbf{97}, 023705\ (2005).
\bibitem{Rep} D. B. A. Rep, A. F. Morpurgo and T. M. Klapwijk,
Org. Elec. \textbf{4}, 201\ (2003).
\bibitem{Hoshino} S. Hoshino, M. Yoshida, S. Uemura, T. Kodzasa, N. Takada, T. Kamata, and K. Yase, J. Appl. Phys. \textbf{95}, 5088 (2004).
\bibitem{Aldrich} Sigma-Aldrich Inc., St. Louis, MO.
\bibitem{Gau} N. J. Watkins, L. Yan and Y. Gao, Appl. Phys. Lett.
\textbf{80}, 4384\ (2002).
\bibitem{Kahn} W. Gau and A. Kahn, Appl. Phys. Lett. \textbf{79},
4040\ (2001).
\bibitem{Wan} A. Wan, J. Hwang, F. Amy, and A. Kahn, Org. Elec.
\textbf{6}, 47\ (2005).


\end{thebibliography}
\end{document}